\pdfoutput=1
\documentclass[a4paper]{article}
\usepackage[OT1]{fontenc}
\usepackage{hyperref}
\hypersetup{
  colorlinks,
  linkcolor={black},
  citecolor={black},
  urlcolor={black}
}

\usepackage{booktabs}
\usepackage{microtype}
\usepackage{spconf,amsmath,graphicx}
\usepackage[table]{xcolor}

\usepackage{csquotes}
\usepackage{siunitx}
\providecommand{\qty}[2]{\SI{#1}{#2}}

\usepackage{fig/targetpitch}

\providecommand\citep{\cite}
\providecommand\citet{\cite}

\usepackage{math}
\usepackage{cleveref}

\newcommand\demopage{http://recherche.ircam.fr/anasyn/bous/vasab2024}
\newcommand\vasb{VaSAB}

\newcommand\dssp{\textsf{sp}}
\newcommand\dssi{\textsf{si}}
\newcommand\dsvo{\textsf{vo}}

\newcommand\rabo{\textsf{RaBo}}
\newcommand\hibo{\textsf{HiBo}}
\newcommand\globo{\textsf{GloBo}}
\newcommand\nobo{\textsf{NoBo}}

\newcommand\Href{H.ref.}
\newcommand\world{WORLD}
\newcommand\spN{\textsf{Nosp}}
\newcommand\siN{\textsf{Nosi}}
\newcommand\spR{\textsf{Rasp}}
\newcommand\siR{\textsf{Rasi}}
\newcommand\voH{\textsf{Hivo}}
\newcommand\voR{\textsf{Ravo}}

\newcommand\best[1]{{\cellcolor[gray]{0.8}} #1}

\newlength\cramfloat
\newlength\cram
\title{%
  \vasb:
  The variable size adaptive information bottleneck
  for disentanglement on speech and singing voice
  }
\name{
  \href{mailto:frederik.bous@ircam.fr}{Frederik Bous},
  Axel Roebel%
  \thanks{%
    This work has been funded partly by the ANR project ARS (ANR-19-CE38-0001-01).
    This work was performed using HPC resources from GENCI-IDRIS
    (Grants 2020-AD011011378 and 2021-AD011011177).
  }
}
\address{
  Science and Technology of Music and Sound (STMS) \\
  IRCAM, Sorbonne Universit\'e, CNRS, Minist\`ere de la Culture \\
}

\begin{document}
  \ninept
  \maketitle
  \begin{abstract}

    The information bottleneck auto-encoder
    is a tool for disentanglement
    commonly used for voice transformation.
    The successful disentanglement
    relies on the right choice of bottleneck size.
    Previous bottleneck auto-encoders created the bottleneck
    by the dimension of the latent space
    or through vector quantization
    and had no means to change the bottleneck size
    of a specific model.
    As the bottleneck removes information
    from the disentangled representation,
    the choice of bottleneck size
    is a trade-off between disentanglement
    and synthesis quality.
    We propose to build the information bottleneck using dropout
    which allows us to change the bottleneck through the dropout rate
    and investigate adapting the bottleneck size depending on the context.
    We experimentally explore
    into using the adaptive bottleneck for pitch transformation
    and demonstrate that the adaptive bottleneck
    leads to improved disentanglement of the F0 parameter for both,
    speech and singing voice leading to improved synthesis quality.
    Using the variable bottleneck size,
    we were able to achieve disentanglement
    for singing voice including extremely high pitches and
    create a universal voice model,
    that works on both speech and singing voice
    with improved synthesis quality.

  \end{abstract}
  \begin{keywords}
    Bottleneck auto-encoder, voice conversion, singing voice transformation
  \end{keywords}
  \section{Introduction}
  \label{sec:intro}

  The goal of voice transformation
  is to change some voice attributes
  in existing voice recordings
  while preserving or adapting other voice attributes
  according to the change.
  Among the changeable voice attributes
  we find research concerned with
  voice identity \citep{qian2019autovc,choi2023nansy,kovela2023any},
  emotion \citep{zhou2022emotional,lemoine2021towards},
  pitch \citep{choi2023nansy,bous2022bottleneck},
  vocal roughness \citep{gentilucci2019composing},
  voice level \citep{molina2014parametric,perrotin2016vocal,bous2023analysis},
  and many more.

  During the age of classical parametric vocoders,
  the lower level voice attributes, like the pitch or voice level,
  could be transformed by adjusting the vocoder parameters
  according to the desired change of the voice attributes
  using predetermined heuristics
  \citep{molina2014parametric,perrotin2016vocal,gentilucci2019composing}.
  To create realistic transformations, however,
  even changing the pitch is not as straightforward
  as simply changing the $f_0$ parameter
  \citep{farner2009natural,degottex2013mixed,ardaillon2017synthesis}:
  Because parameters of real voice signals are not independent,
  a change in one parameter
  needs to be accompanied by a change
  in the other parameters.

  With deep learning,
  modelling the dependencies
  between vocoder parameters and voice attributes has become feasible
  and has lead to significant improvements in naturalness.
  To model the dependencies,
  disentanglement can be used,
  to create a parametrization of independent parameters.
  Disentanglement has been applied to individual attributes, like
  voice identity \citep{qian2019autovc} or pitch \citep{bous2022bottleneck}
  and has been used to disentangle multiple parameters simultaneously
  \citep{qian2020unsupervised,choi2023nansy,bous2023analysis}.

  Different approaches to disentanglement can be found in the literature,
  but most approaches rely on auto-encoders
  to create the disentangled reparametrization
  which then consists of the control parameters
  and the latent codes
  of one or multiple encoders.
  Independence of the different parameter groups
  can be achieved through adversarial approaches \citep{lample2017fader,chou2018multi},
  through information bottlenecks \citep{qian2019autovc,wu2020one}
  or using inductive biases like instance normalization \citep{chou2019one}
  or perturbation \citep{qian2020unsupervised,choi2023nansy}.
  An information bottleneck has been achieved
  through dimensionality reduction and sub-sampling \citep{qian2019autovc}
  or vector quantization \citep{wu2020one}.

  While the information bottleneck is a powerful tool for disentanglement,
  it has lost in popularity in recent years
  as the bottleneck tends to remove information
  not only related to the disentangled attribute \citep{choi2021neural}.
  To address these issues
  we propose a new bottleneck type
  which can be widened and narrowed
  depending on the underlying data
  and apply the proposed bottleneck
  to the task of transposition.
  \vspace*{\cram}

  \subsection{The bottleneck auto-encoder}
  \label{ssec:bottleneck}

    \newcommand\dimin{n_\mathrm{in}}
    \newcommand\dimbn{n_b}
    \newcommand\dimla{n_l}
    \newcommand\pfull{p_g}
    The bottleneck auto-encoder has been introduced
    as AutoVC for voice identity conversion \citep{qian2019autovc}.
    Formally an auto-encoder consists of an encoder $E$ and a decoder $D$.
    The encoder $E$ maps its input $x\in\reals^{\dimin{}}$
    to the latent code $c\in\reals^{\dimla{}}$.
    The decoder $D$ maps the latent code $c$
    together with some control input $y$
    back to the original space,
    trying to approximate $x$ by $D(c, y) = \hat{x} \in\reals^{\dimin{}}$.
    Furthermore, additional perturbations on the latent code
    can be applied on the latent code
    to produce a stronger bottleneck
    like temporal sub-sampling \citep{qian2019autovc}
    or vector quantization \citep{wu2020one}.

    Using a voice attribute $a$ as the control input $y$
    allows disentangling the latent code $c$ from the attribute $a$.
    The fundamental idea of the information bottleneck auto-encoder
    is that the latent code $c$ cannot fully express the input $x$.
    Since the decoder $D$ has direct access to $a$ through $y$,
    the encoder will remove information related to $a$
    from the latent code $c$.
    A successful application of the bottleneck auto-encoder
    relies on carefully choosing the channel capacity of the bottleneck.
    If the bottleneck is too large,
    the control parameter may not be fully removed
    from the latent code of the auto-encoder
    and transformation of the control parameter is not possible.
    If the bottleneck is too narrow,
    relevant information is removed in addition to the control parameter
    and the quality of the resynthesis is degraded.
    The optimal bottleneck size,
    however,
    depends on the underlying data
    and might vary depending on many factors.
    In previous work
    this bottleneck was hard-wired into the architecture of the auto-encoder.
    The bottleneck size was fixed and could not be altered.
    However, as observed in \citet{bous2022bottleneck},
    different voice types required different optimal bottleneck sizes.
    As a consequence, using a fixed-size bottleneck auto-encoder
    does not allow for creating universal voice models.
    In this paper we propose an alternative way
    to create an information bottleneck based on dropout,
    which allows adapting the bottleneck size as required.
    \vspace*{\cram}

    \section{Proposed method}
    \label{ssec:vasb}

    In this section we introduce
    the \textbf{Va}riable \textbf{S}ize \textbf{A}daptive information \textbf{B}ottleneck (\vasb)
    which limits the information flow from encoder to decoder
    using dropout on the latent code.
    Suppose a latent space of dimension $\dimla{}$
    allows a channel capacity of $c$.
    Then, dropout of rate $r$
    will reduce that channel capacity
    by the same factor $r$ to $c - rc = (1-r) c$.
    Thus, an effective bottleneck of size $\dimbn{}$
    can be achieved by a dropout rate of $r = 1 - \dimbn{}/\dimla{}$.
    In the following we will introduce three different dropout types
    that can be used with \vasb.
    \vspace*{\cram}

    \subsection{\textbf{Ra}ndom dropout \textbf{bo}ttleneck: \rabo}

    \rabo\ uses (regular) dropout on each feature independently.
    The dropout rate $r$ can be varied over time (or space).
    This allows us to make the bottleneck tighter
    for some data, while widening it for other,
    depending on how much information is contained in the underlying data.
    Since $r$ can take any value between $0$ and $1$,
    the effective bottleneck size can be fractional.
    \vspace*{\cram}

    \subsection{\textbf{Hi}erarchical dropout \textbf{bo}ttleneck: \hibo}

    Inspired by Soundstream \citep{zeghidour2021soundstream}
    we propose an alternative dropout method
    which we call \emph{hierarchical dropout}.
    For \hibo, for each time step
    we first randomly determine how many features will be set to zero
    by sampling from a binomial distribution $\mathcal{B}(\dimla{}, r)$.
    Then, we set the decided amount to zero in a fixed order.
    This has been used in \citet{zeghidour2021soundstream}
    to allow for variable compression rate
    for audio coding,
    and it might allow for more efficient usage
    of the latent code in voice transformation as well.
    \vspace*{\cram}

    \subsection{\textbf{Glo}bal dropout \textbf{bo}ttleneck: \globo}
    \label{ssec:globo}

    The ability to regulate the information bottleneck at will,
    furthermore allows us to handle the bottleneck differently during inference.
    However, opening up the bottleneck completely during inference
    is not trivially possible with the approaches described above alone,
    due to the following problem:
    Using the dropout-based bottleneck,
    the actual latent space can be chosen arbitrarily large.
    With increased $\dimla{}$, the dropout rate $r$ increases.
    Even for moderately large $\dimla{}$
    it becomes rather unlikely
    that during training
    the decoder will see the full latent space%
    \footnote{
      If, for example we wanted to set $\dimla{} = 64$,
      and we were working on singing voice,
      which requires us to use $\dimbn{} = 3$ on average
      as observed in \citet{bous2022bottleneck},
      we would need a dropout rate of $1 - 3 / 64 \approx 95\%$.
      In that case, the probability that the full latent space
      is preserved equals $(3 / 64)^{64} \approx \num{9e-86}$
      which means that for any practical purposes,
      this will never happen.
    }
    and the decoder will learn to expect
    a partially degraded latent space.
    As a consequence,
    the auto-encoder would not learn to utilize the full latent code,
    and we would have to apply dropout during inference as well
    and not be able to use the full capacity of the latent code.
    This effect was observed during our preliminary experiments.

    Thus, we propose to additionally use
    an instance-wise global dropout bottleneck:
    For \globo\ the dropout-rate is averaged
    over the entire training sample.
    Then according to this averaged dropout probability
    the whole latent code of the sample
    is either left untouched,
    or set to zero entirely.
    Setting the entire latent code sometimes to zero
    furthermore incentivizes the model to use its control input $y$,
    as $y$ is the only input
    that will reliably be available
    at all times.

    While \globo\ itself is not sufficient
    to achieve any meaningful disentanglement,
    it can be combined with \rabo\ and \hibo,
    to encourage the auto-encoder
    to use the full latent code during inference.
    During training,
    with probability $\pfull{}$,
    we apply \globo\ to a sample
    instead of \rabo\ or \hibo.
    \vspace*{\cram}

    \subsection{Combining speech and singing voice}
    \label{ssec:speechsing}

    As shown in \citet{bous2022bottleneck},
    speech and singing voice require different bottleneck capacities.
    Accordingly, training a single model for both speech and singing voice
    was not possible with the approach proposed in \citet{bous2022bottleneck}.
    The adaptive control of the target bottleneck capacity
    allows changing the bottleneck capacity
    depending on the voice type of the training samples.
    As will be demonstrated in the experimental evaluation,
    the proposed approach does now allow training a model
    that can successfully work with both,
    singing and spoken voice.
    \vspace*{\cram}

    \section{Experiments}
    \label{sec:experiments}

    We have developed \vasb\
    in order to improve our pitch-transformation
    from \citet{bous2022bottleneck}.
    Therefore, we will perform the experimental evaluation for \vasb\
    on the task of pitch-transformation.
    We evaluate \vasb\ on two studies:
    First, we will explore the mean \fo\ error
    as a function over the target \fo\
    generated by different training configurations.
    This will allow us understanding how different hyperparameters
    influence the disentanglement ability of \vasb.
    Second, we perform a perceptual test
    to evaluate the synthesis quality of selected models
    and to compare them to other pitch-transformation methods,
    in particular the fixed-size bottleneck.
    \vspace*{\cram}

    \subsection{Training data}
    \label{ssec:dataset}

    We will use three different training datasets:
    a pure singing voice dataset \dssi,
    a pure speech dataset \dssp,
    and a combined dataset \dsvo\ consisting of both singing and speech.
    We use the same singing voice dataset
    as in \citet{bous2022bottleneck} and \citet{bous2023analysis},
    which consists of the datasets from 
    \citet{
      tsirulnik2019singing,
      duan2013nus,
      grammalidis2016treasures,
      koguchi2020pjs,
      tamaru2020jvs,
      ogawa2021tohoku,
      wilkins2018vocalset,
      ardaillon2017synthesis,
      deschamps2022bmue,
      bous2023neural%
    }.
    The speech dataset consists of VCTK \citep{yamagishi2019vctk},
    Att-HACK \citep{lemoine2020atthack},
    and further internal datasets containing expressive speech,
    amounting to about \qty{70}{\hour} of speech in total.

    Training data for the composed datasets \dsvo\
    is sampled 50/50 from \dssp\ and \dssi,
    such that both speech and singing
    occur with equal probability.
    Furthermore, as observed in \citet{bous2023neural}
    (e.g.\ fig.\ 2.12),
    high-pitched singing voice was actually extremely rare
    in the dataset.
    Therefore, within \dssi, we employed a sampling strategy
    that would draw recordings with pitches above \qty{700}{\hertz}
    with a probability of about 10\%.
    \vspace*{\cram}

    \begin{figure*}
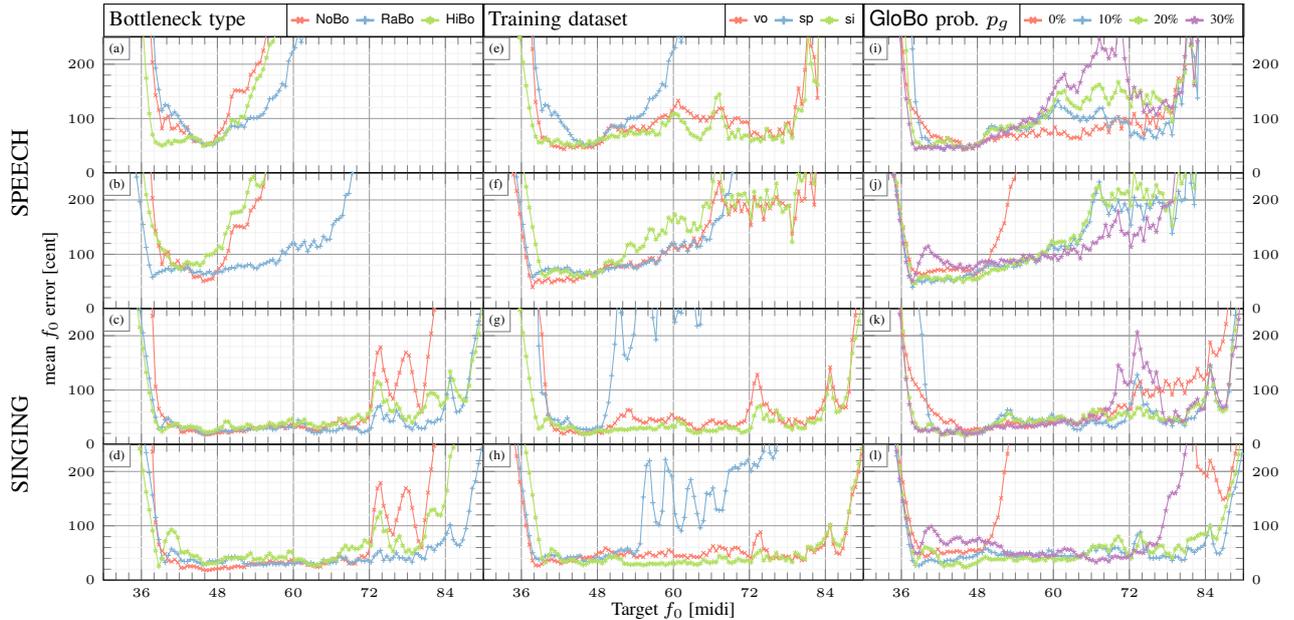

      \centering
      \plotAllTargetPitch
      \par\vspace*{-5mm}
      \caption[Mean \fo-error  a function of target \fo.]{%
        Mean \fo-error [cent] a function of target \fo\ [midi].
        The top two rows were evaluated on speech,
        the bottom two rows were evaluated on singing.
        Each column varies a different hyperparameter
        inside each subplot.
        The models in the first column,
        are trained on the voice-type specific datasets
        ((a) and (b) on speech, (c) and (d) on singing)
        and with $\pfull{}=10\%$.
        The models in the middle column
        are trained using \rabo\ with $\pfull{}=10\%$.
        The models in the right column
        are trained on the voice dataset
        using \rabo.
        The latent code sizes $\dimla{}$ are 16
        for the odd rows ((a), (c), (e), (g), (i) and (k))
        and 64 for the even rows
        ((b), (d), (f), (h), (j) and (l)).
      }
      \vspace*{\cramfloat}
      \label{fig:f0acc}
    \end{figure*}

    \subsection{Model architecture}
    We use the same architecture
    and the training procedure described in \citet{bous2022bottleneck}
    with the only difference that
    instead of conditioning both
    the encoder and decoder blocks of the auto-encoder,
    here we condition only the decoder with the target \fo.
    This change is motivated by the fact
    that the conditioning of the encoder
    did not have a positive effect in the evaluation,
    but has a negative effect during inference
    as we have to estimate the \fo\ of the input signal
    to be able to use the  model for voice transformation.
    \vspace*{\cram}

    \subsection{\fo\ estimation}

    To evaluate the \fo,
    we train an \fo-estimator
    using the annotation method from \citet{ardaillon2019fully}.
    However, here the \fo-estimator uses the mel-spectrogram as input
    instead of the raw-audio.
    The \fo-estimator was trained on a resynthesis of
    \dsvo\ from \cref{ssec:dataset}
    using the \world vocoder \citep{morise2016world}
    and had a final validation error of \qty{2.27}{\hertz} and \qty{22.2}{cent}.
    Using a pitch estimator based on the mel-spectrogram
    allows us to evaluate the output of the auto-encoder
    without relying on additional tools
    to convert the mel-spectrogram back to audio,
    thus observing the performance of the auto-encoder
    in isolation.
    \vspace*{\cram}

    \subsection{Audio synthesis}

    To convert the mel-spectrograms to audio
    for the perceptual test,
    we use the neural vocoder of \citet{roebel2022neural}
    which has been shown to work particularly well
    on singing voice.
    We use the same universal voice model
    (trained on speech and singing voice)
    for synthesis of both speech and singing voice.
    \vspace*{\cram}

    \subsection{Models}

    Models are varied in their bottleneck type
    (\nobo: no dropout,
     \rabo: random dropout,
     \hibo: hierarchical dropout)
    training dataset (\dssp: speech, \dssi: singing, \dsvo: voice (\dssp\ + \dssi)),
    latent code size $\dimla{}$ (16 and 64)
    and \globo\ probability $\pfull{}$ (0.0, 0.1, 0.2, 0.3, see \cref{ssec:globo}).
    According to the findings of \citet{bous2022bottleneck},
    and in accordance with preliminary experimenting with \rabo\ and \hibo,
    we set $\dimbn{}$ to 3 for singing voice and to 8 for speech,
    and $\dimbn{} = \dimla{}$ for any unvoiced frame.

    In the perceptual test we compare the following models:
    \begin{description}
      \itemsep-0.3mm
      \item[\spN] Baseline for speech from \citet{bous2022bottleneck} with $\dimla{}=8$
      \item[\siN] Baseline for singing voice from \citet{bous2022bottleneck} with $\dimla{}=3$
      \item[\spR] \rabo\ model with $\dimla{} = 64$, $\pfull{} = 0.1$ trained on \dssp
      \item[\siR] \rabo\ model with $\dimla{} = 64$, $\pfull{} = 0.1$ trained on \dssi
      \item[\voR] \rabo\ model with $\dimla{} = 64$, $\pfull{} = 0.1$ trained on \dsvo
      \item[\voH] \hibo\ model with $\dimla{} = 64$, $\pfull{} = 0.1$ trained on \dsvo
      \item[\world] Signal processing baseline using the \world\ vocoder~\citep{morise2016world}
      \item[\Href] Original recordings included as hidden reference
    \end{description}
    Audio samples can be found on our demo website.\footnote{\url{\demopage}}
    \vspace*{\cram}

    \section{Results}
    \label{sec:results}
    \subsection{\fo\ accuracy}
    \label{ssec:accuracy}

    In this first experiment we investigate
    the effect of different hyperparameters
    on the disentanglement capabilities of the models.
    We conduct our investigation by examining
    the mean \fo\ error as a function of the target \fo\
    in a transposition task.
    The curves are plotted in \cref{fig:f0acc}.
    This relationship gives us insight
    to where the \fo\ is well disentangled
    and where problems occur:
    For example,
    the valleys in the \fo\ error in \cref{fig:f0acc} (c) and (d)
    for target \fo\ above C5 (midi~72; \qty{523}{\hertz})
    occur when a model fails to produce continuous \fo-curves
    and starts to discretize the \fo.
    In such cases, additionally to being off-key,
    the normally continuously changing \fo\
    jumps abruptly between different values
    and creates very unnatural behaving voice.
    This phenomenon can be found in all models
    to a certain extent,
    but is by far most apparent for \nobo.
    Similarly, isolated peaks,
    as, e.g.\ in \cref{fig:f0acc} (g) and (h)
    around C\#5 (midi~73; \qty{554}{\hertz})
    and C\#6 (midi~85; \qty{1190}{\hertz}),
    suggest that the models
    have difficulties to synthesize this particular pitch
    and tend to systematically be out of tune
    for that particular pitch.
    Note that, in general however,
    the \fo, error is not necessarily the same
    as the error in perceived pitch
    and is caused by effects like
    imperfect timing during changing pitches
    (which is why the \fo\ error is significantly larger for speech)
    rather than actually being out of tune.
    \vspace*{\cram}

    \subsubsection{Effect of bottleneck type}

    In the first column of \cref{fig:f0acc}
    ((a), (b), (c) and (d))
    we compare the different bottleneck types.
    In all cases the pitch range is substantially extended
    by \rabo\ relative to \nobo.
    \hibo\ improves disentanglement for singing
    and improves disentanglement for speech
    in configuration (a) but not (b).
    \rabo\ generally outperforms \hibo.

    For singing (\cref{fig:f0acc} (c) and (d)), furthermore, we can see
    the typical problem of \nobo,
    when applied to high pitches,
    which is that \nobo\ is not able
    to produce a continuous range of pitches
    above a certain threshold (here C5; midi~72; \qty{523}{\hertz}).
    In this range, the \nobo\ model becomes biased towards
    only a few pitches (midi~72; 76 and 80)
    and produces very unnatural sounding results.
    Similarly, the accuracy of \hibo\
    decreases sharply above C5,
    though the discretization effect is not so extreme,
    and the model can still somewhat produce fluent pitch glides.

    The only case where \hibo\ is clearly better
    than all other models
    is the lower \fo\ values for speech
    on the $\dimla{}=16$ version (a).
    Note, however that this is still outperformed
    by the \rabo\ model with $\dimla{}=64$ ((b)).
    \vspace*{\cram}

    \subsubsection{Effect of training data}

    The middle column of \cref{fig:f0acc}
    ((e), (f), (g) and (h))
    visualizes the influence of the voice type in the training data
    on the disentanglement.
    When transforming speech
    (\cref{fig:f0acc} (e) and (f)), we can see that
    having singing voice in the dataset
    allows transforming to much higher pitches
    than with a pure speech dataset alone.
    The best disentanglement is achieved,
    when trained on both speech and singing voice.

    When transforming singing voice
    (\cref{fig:f0acc} (g) and (h)),
    the presence of speech in the training data
    seems not to have a big influence on the result
    (as long as it is still also trained on singing).
    However, the pure singing model
    has slightly less difficulty around
    C\#5 (midi~73; \qty{554}{\hertz}),
    especially for $\dimla{}=16$.
    For lower pitches, the specialized singing model
    allows transposing lower
    in the case of $\dimla{}=16$ (g),
    whereas the general voice model
    allows lower \fo's
    in the case of $\dimla{}=64$ (h),
    where the highest pitch range
    of all models is achieved.
    \vspace*{\cram}

    \subsubsection{Effect of global dropout probability}

    Finally, we'll have a look at the effect
    of \globo\ as described in \cref{ssec:globo}
    using the graphs of \cref{fig:f0acc} (i), (j), (k) and (l).
    The most clear observation is that,
    as predicted,
    for the large latent space ($\dimla{}=64$, (j) and (l))
    it is absolutely necessary to apply global dropout for some samples.
    In this case, the optimal value for $\pfull{}$ seems to be
    somewhere between 0.1 and 0.2.
    A too large $\pfull{}$, however, seems to be rather detrimental.
    For the small latent space ($\dimla{}=16$, (i) and (k))
    the global dropout is still useful,
    but its effect seems to be less severe.
    Again, a too large $\pfull{}$ affects
    the disentanglement negatively.
    \vspace*{\cram}

    \subsection{Synthesis quality}

    Participants were asked in an online survey
    where they had to rate the naturalness of given audio samples
    on a scale from 1 to 5.
    We conduct separate tests for speech and singing voice
    and received 50 responses for each test.

    The results of the perceptual test are shown in \cref{tab:mos:speech,tab:mos:singing}.
    Generally we can observe that
    among the information bottleneck approaches
    the rigid bottleneck, \spN\ and \siN,
    is surpassed by all other proposed bottleneck types in most situations
    and has never obtained the best score in any situation.
    This shows that \vasb\ provides an improvement in performance.
    Similarly, the signal processing baseline, \world,
    is consistently outperformed by at least one \vasb\ models
    (though not always the same).

    For speech, no model can be declared as superior in general
    as different models perform better than others in different situations.
    It is interesting to note that even though training with singing voice included
    higher pitches can be synthesized,
    but their naturalness is rather lacking.
    This is not surprising,
    as extremely high speaking voice is rather rare
    and thus may provoke participants to think it is unnatural.

    For singing voice, the model \voR\ emerges as the best
    as it has received the best scores in 4 out of 5 categories.
    For strong upwards transpositions \voR\ performs not so well
    and is surpassed by \world\ and \siR.
    These strong transpositions up are difficult edge cases,
    and it is plausible that the specialized singing model
    is more adapted in this unique situation.

    \begin{table}
      \resizebox{\linewidth}{!}{
        \small
        \begin{tabular}{lccccc}
          \toprule
          Transp.\ [cent] & $-1600$ & $-800$ & $0$ & $800$ & $1600$ \\
          \midrule
          \Href & & & $4.49 \pm 0.13$ & & \\
\world\ \citep{morise2016world} & $1.67 \pm 0.30$ & $2.67 \pm 0.37$ & $3.72 \pm 0.44$ & $2.64 \pm 0.38$ & $2.57 \pm 0.40$\\
\spN\ \citep{bous2022bottleneck} & $1.78 \pm 0.33$ & $2.61 \pm 0.39$ & $3.67 \pm 0.47$ & $2.86 \pm 0.33$ & $1.47 \pm 0.27$\\
\spR & \best{$2.25 \pm 0.39$} & $2.89 \pm 0.44$ & $3.86 \pm 0.34$ & $3.00 \pm 0.44$ & \best{$2.75 \pm 0.46$}\\
\voH & $2.06 \pm 0.36$ & \best{$3.06 \pm 0.45$} & \best{$3.94 \pm 0.37$} & $3.06 \pm 0.36$ & $2.11 \pm 0.40$\\
\voR & \best{$2.25 \pm 0.37$} & $2.75 \pm 0.42$ & $3.50 \pm 0.42$ & \best{$3.08 \pm 0.39$} & $2.29 \pm 0.39$

        \end{tabular}
      }
      \vspace{-2mm}
      \caption{Mean opinion score for speech.}
      \vspace{1mm}
      \label{tab:mos:speech}
      \resizebox{\linewidth}{!}{
        \small
        \begin{tabular}{lccccc}
          \toprule
          Transp.\ [cent] & $-1600$ & $-800$ & $0$ & $800$ & $1600$ \\
          \midrule
          \Href  & & & $4.52 \pm 0.12$ & & \\
\world\ \citep{morise2016world} & $1.33 \pm 0.20$        & $2.48 \pm 0.44$        & $3.97 \pm 0.35$        & $3.13 \pm 0.41$ & $2.57 \pm 0.48$\\
\siN\ \citep{bous2022bottleneck}  & $2.00 \pm 0.41$        & $2.63 \pm 0.47$        & $4.03 \pm 0.39$        & $2.53 \pm 0.51$ & $1.87 \pm 0.45$\\
\siR   & $2.57 \pm 0.46$        & $3.23 \pm 0.42$        & $3.83 \pm 0.37$        & $3.35 \pm 0.44$ & \best{$3.42 \pm 0.42$}\\
\voH   & $2.16 \pm 0.48$        & $3.26 \pm 0.39$        & $4.06 \pm 0.34$        & $3.45 \pm 0.38$ & $1.84 \pm 0.36$\\
\voR   & \best{$3.00 \pm 0.48$} & \best{$3.29 \pm 0.44$} & \best{$4.17 \pm 0.35$} & \best{$3.65 \pm 0.47$} & $2.50 \pm 0.44$

        \end{tabular}
      }
      \vspace{-2mm}
      \caption{Mean opinion score for singing.}
      \vspace*{\cramfloat}
      \label{tab:mos:singing}
    \end{table}

    As seen in \cref{ssec:accuracy}
    \hibo\ does not achieve as much disentanglement
    as \rabo.
    Since the ordered dropout is more predictable
    it might not create such a narrow bottleneck
    as the random dropout with the same rate.
    Thus, our model with the hierarchical dropout
    might be more reliable in areas where disentanglement is easy
    and fail in the more special cases,
    which might explain why it performs better
    for smaller transpositions.
    This does not necessarily mean
    that \hibo\ is less suited for \vasb,
    since the right bottleneck size is crucial
    for successful transformations.
    A study concerned with finding the optimal dropout rate
    for \hibo\ might allow us to explore
    the full potential of \hibo.
    \vspace*{\cram}

    \section{Conclusion}
    \label{sec:conclusion}

    In this paper we introduced the
    \textbf{Va}riable \textbf{S}ize \textbf{A}daptive information \textbf{B}ottleneck (\vasb)
    as a method to enforce disentanglement on the latent code
    of an information bottleneck auto-encoder.
    \vasb\ works by applying dropout to the latent code
    and can therefore be adaptively changed, depending on the underlying data.
    We have applied \vasb\ to the task of transposition of human voice
    and were able to create a universal voice model
    which works on both speech and singing voice.
    With previous non-adaptive bottlenecks
    a simultaneous treatment of speech and singing was not possible
    due to the different information rate of the voice types.

    We investigated the effect of various hyperparameters
    related to \vasb\
    on the disentanglement capabilities on the transposition task.
    The key findings were,
    that \rabo, the bottleneck based on random dropout,
    performs best among the examined bottleneck types.
    Using \vasb\ allows training universal voice models,
    with similar quality than the specialized models,
    in some cases the specialized models were better,
    while in other cases the universal models were better.

    The key advantage of \vasb\ is,
    that it decouples the effective bottleneck size $\dimbn{}$
    from the dimension of the latent code $\dimla{}$.
    Therefore, the bottleneck is not structural
    and can be opened when it is not required.
    This could allow for combining information bottleneck auto-encoders
    with other disentanglement approaches
    which could be explored in future work.

  \bibliographystyle{IEEEbib}
  \bibliography{refs_short}

\end{document}